\documentclass[11pt]{amsart}

\pdfoutput = 1

\usepackage[english]{babel}
\usepackage[autostyle]{csquotes}
\usepackage{amsmath,amssymb,amsthm}
\usepackage{amsfonts}
\usepackage{amsxtra}
\usepackage{xcolor}

\usepackage{array,dcolumn}
\usepackage{graphicx}
\usepackage[hyperfootnotes=true,
        pdffitwindow=true,
        plainpages=false,
        pdfpagelabels=true,
        pdfpagemode=UseOutlines,
        pdfpagelayout=SinglePage,
                hyperindex,]{hyperref}

\usepackage{lmodern}
\usepackage[T1]{fontenc}
\usepackage[utf8]{inputenc}
\usepackage[activate={true,nocompatibility},spacing,kerning]{microtype}
\usepackage{bm}
\usepackage{bbm}
\usepackage[sc,osf]{mathpazo}
\microtypecontext{spacing=nonfrench}

  \calclayout

\newcommand{\mathsym}[1]{{}}
\newcommand{\unicode}[1]{{}}

\theoremstyle{plain}

\theoremstyle{definition}

\theoremstyle{remark}

\numberwithin{equation}{section}

\begin{document}


\title{Computable structural formulas for the distribution of the $\beta$-Jacobi edge eigenvalues}
\author{Peter J. Forrester}
\address{School of Mathematics and Statistics, 
ARC Centre of Excellence for Mathematical \& Statistical Frontiers,
University of Melbourne, Victoria 3010, Australia}
\email{pjforr@unimelb.edu.au}

\author{Santosh Kumar}
\address{
Department of Physics, Shiv Nadar University, Gautam Buddha Nagar, Uttar Pradesh 201314, India
}
 \email{skumar.physics@gmail.com}

\date{\today}


\dedicatory{To the memory of Richard Askey}

\begin{abstract}
The Jacobi ensemble is one of the classical ensembles of random matrix theory.
Prominent in applications are properties of the eigenvalues at the spectrum
edge, specifically the distribution of the largest (e.g.~Roy's largest root test
in multivariate statistics) and smallest (e.g.~condition numbers of linear systems)
eigenvalues. We identify three ranges of parameter values for which the gap
probability determining these distributions is a finite sum with respect to 
particular bases, and moreover make use of a certain differential-difference
system fundamental in the theory of the Selberg integral 
to provide a recursive scheme to compute the corresponding coefficients.
\end{abstract}


\maketitle


    \setcounter{section}{-1}
 \section{Prologue}
 
 The Selberg integral is the $N$-dimensional generalisation of the Euler $\beta$-integral
$$
J_{\lambda_1,\lambda_2,\beta,N} = 
\int_0^1 dx_1 \cdots \int_0^1 dx_N \,
\prod_{l=1}^N x_l^{\lambda_1} (1 - x_l)^{\lambda_2} \prod_{1 \le j < k \le N}
| x_k - x_j |^\beta.
$$
In a paper published in 1944 \cite{Se44} Selberg derived the product of gamma function evaluation, specified by (\ref{1X}) below. As detailed in the review \cite{FW08}, knowledge of this result went almost completely unnoticed for over 30 years, until a number of coincidences led to the realisation that it provides a proof of  the conjecture
\begin{equation}\label{1.6d}
{1 \over (2 \pi)^{n/2} } \int_{-\infty}^\infty \cdots  \int_{-\infty}^\infty 
\prod_{j=1}^n e^{- t_i^2/2} \prod_{1 \le j < k \le n} | t_j - t_k|^{2 \gamma}
\, dt_1 \cdots dt_n = \prod_{j=1}^n {\Gamma(1 + j \gamma) \over \Gamma(1 + \gamma)},
\quad {\rm Re} \, \gamma > -1/n,
\end{equation}
formulated by Mehta and Dyson \cite{MD63} in the context of random matrix theory applied to nuclear physics.

Outside of the original paper, the now ex-conjecture (\ref{1.6d}) appeared in the first edition of Mehta's book \cite{Me67} and also, in 1974, in the problem section of SIAM review \cite{Me74}. From the source \cite{As98}, we learn that it was from this latter publication that Askey took an active interest in this class of multiple integrals. At the beginning of the '80's he was responsible for initiating the study of $q$-generalisations \cite{As80}, and also supervised the well known PhD thesis of Morris \cite{Mo82} on the wider theory, by then understood to be related to root systems from the theory of Lie algebras.

In 1987 Aomoto's \cite{Ao87} work on proving (and extending) Selberg's result by deriving the recurrence in $\lambda_1$
$$
J_{\lambda_1+1,\lambda_2,\beta,N} = \prod_{p=0}^N {\lambda_1 + 1 + p \beta/2 \over
\lambda_1 + \lambda_2 + 2 + (N-1+p)\beta/2} \, J_{\lambda_1,\lambda_2,\beta,N}
$$
was published. This work, which was underpinned by a sophisticated viewpoint of the theory  of multidimensional hypergeometric functions relating to de Rham cohomology, later summarised in the book \cite{AK11}, makes use of nothing beyond skilful application of integration by parts.  

One of us (PJF) had the opportunity to visit Askey in Madison for two months from late August in 1988. In discussions, it came up how fundamental Askey viewed the recurrence relation viewpoint on the gamma function evaluation of the Euler integral --- the case $N =1$ of (\ref{1X}), and how he often used it as extension material in a gifted high school student program as an example of powerful ideas which get missed in conventional undergraduate syllabi.  However, as expressed in his 1975 book \cite{As75}, Askey's broader program was "to study special functions not for their own sake, but to be able to use them to solve problems". In relation to the Selberg integral, a number of its uses and consequences to topics such as random matrix theory, conformal field theory, and statistical properties of the zeros of the Riemann zeta function are reviewed in \cite{FW08}. Subsequently the Selberg integral has facilitated a number recent advances in the field of Gaussian multiplicative chaos \cite{Os18}. 

In the present work we return to the recursive structures of the type introduced in \cite{Ao87}, and contribute to Askey's broader program by identifying certain of their applications within random matrix theory.

\section{Introduction}\label{S1}

\subsection{The Jacobi ensemble}
In random matrix theory the Jacobi (or more precisely $\beta$--Jacobi) ensemble refers to the
class of eigenvalue probability density functions
\begin{equation}\label{1}
{\mathcal P}^{\rm J}(x_1,\dots,x_N) : = {1 \over J_{\lambda_1,\lambda_2,\beta,N}}
\prod_{l=1}^N x_l^{\lambda_1} (1 - x_l)^{\lambda_2} \chi_{x_l \in (0,1)} \prod_{1 \le j < k \le N} | x_k - x_j|^\beta.
\end{equation}
Here $\chi_J = 1$ for $J$ true, $\chi_J = 0$ otherwise, and $J_{\lambda_1,\lambda_2,\beta,N}$ is the normalisation ---
it is the Selberg integral (see e.g.~\cite[Ch.~4]{Fo10}), given by
\begin{equation}\label{1X}
J_{\lambda_1,\lambda_2,\beta,N}=\prod_{j=0}^{N-1}\frac{\Gamma(\lambda_1+1+j\beta/2)\Gamma(\lambda_2+1+j\beta/2)\Gamma(1+(j+1)\beta/2)}{\Gamma(\lambda_1+\lambda_2+2+(N+j-1)\beta/2)\Gamma(1+\beta/2)}.
\end{equation}
An example of the Jacobi ensemble --- although not then named as such --- first revealed itself in studies of
multivariate statistics; see e..g.~\cite{An58,Mu82}. Specifically, let $X_i$ $(i=1,2)$ be $n_i \times N$ $(n_i \ge N)$
standard real Gaussian matrices, to be thought of as centred random data matrices. Let $W_i = X_i^T X_i$ $(i=1,2)$
denote the corresponding $N \times N$ covariance matrices. One has that the eigenvalues of
\begin{equation}\label{2}
(\mathbb I + W_1^{-1} W_2)^{-1}
\end{equation}
form the Jacobi ensemble with parameters
\begin{equation}\label{3}
(\lambda_1, \lambda _2, \beta) = \Big ( {1 \over 2}(n_1 - N - 1), {1 \over 2}(n_2 - N - 1),1 \Big ).
\end{equation}

This result can be generalised to the cases that the matrices $X_i$ contain standard complex Gaussian entries,
or standard quaternion Gaussian entries, with the latter represented in $X_i$ as $2 \times 2$ complex blocks
of the form
$$
\begin{bmatrix} z & w \\ - \bar{w} & z \end{bmatrix}, \qquad w,z \in \mathbb C.
$$
One has (see \cite[Prop.~3.6.1]{Fo10}) that the eigenvalue probability density function of
(\ref{2}) is now given the by the Jacobi ensemble with parameters
\begin{equation}\label{4}
(\lambda_1, \lambda _2, \beta) = \Big ( {\beta \over 2}(n_1 - N + 1 - 2/\beta), {\beta \over 2}(n_2 - N + 1 - 2/\beta),\beta \Big ),
\end{equation}
where $\beta = 2$ (complex entries) and $\beta = 4$ (quaternion entries). Note from (\ref{3}) that setting $\beta = 1$
in (\ref{4}) specifies the case of real entries in (\ref{2}).

Square standard Gaussian random matrices $X$ can be used to construct Haar distributed unitary matrices according to
\begin{equation}\label{5}
S = X (X^\dagger X)^{-1/2};
\end{equation}
see e..g.~\cite{DF17}. Let $X$ (and thus $S$) be of size $L \times L$, and let $S_{n,N}$ $(n \ge N)$ denote the top
$n \times N$ sub-block of $S$ (since $S$ is Haar distributed, any $n$ rows and $N$ columns will do). Then, as
a corollary of knowledge of the eigenvalues probability density function (\ref{2}), it is possible to make
use of (\ref{5}) to show \cite{Co05,Fo06a} that for $L \ge n + N$ the square singular values of $S_{n,N}$ (i.e.~the eigenvalues
of $S_{n,N}^\dagger S_{n,N}$) have density function given by the Jacobi ensemble with
$$
(\lambda_1, \lambda _2, \beta) = \Big ( {\beta \over 2}(n - N + 1 - 2/\beta), {\beta \over 2}(L - n - N + 1 - 2/\beta),\beta \Big ).
$$
Again $\beta = 1,2$ or 4 according to the entries of $X$ (and thus $S$) being real, complex or quaternion respectively.

Associated with the block decomposition of a real orthogonal matrix $S$ is the so-called cosine-sine (CS) expansion.
Edelman and Sutton \cite{ES06a} showed that a sequence of Householder transformations can be applied to
$S$ to obtain a real orthogonal matrix $S$ which is block bi-diagonal and has the same CS values. The top left
diagonal matrix has the form
\begin{equation}\label{5.1x}
\left [ \begin{array}{cccc} c_N & -s_N c_{N-1}' &&   \\
 & c_{N-1}  s_{N-1}' & \ddots&  \\
   &  & \ddots & -s_2 c_1'   \\
         &       & & c_1 s_1' 
\end{array}
\right ],
\end{equation}
where $c_i = \cos \theta_i$, $s_i = \sin \theta_i$, $c_i' = \cos \phi_i$,
$s_i' = \sin \phi_i$, with all angles between 0 and $\pi/2$. Moreover, with the notation
$B[a,b]$ for the beta distribution, it is shown in \cite{ES06a} that
\begin{eqnarray*}
&& \cos^2 \theta_j \in {\rm B} [ \beta (a + j)/2, \beta (b+j)/2] \qquad
(j=1,\dots,N) \nonumber \\
&& \cos^2 \phi_j \in {\rm B} [ \beta j/2, \beta (a+b+1+j)/2] \qquad
(j=1,\dots,N-1), \nonumber
\end{eqnarray*}
and that the squared singular values of (\ref{5.1x}) have density given by  (\ref{1}) with
$$
(\lambda_1, \lambda _2, \beta) = \Big ( {\beta \over 2}(a+1) - 1 , {\beta \over 2}(b+1) - 1,\beta \Big ).
$$
This construction thus realises the Jacobi ensemble for general parameters.

It is also true that for general parameters the probability density function (1), upon the
change of variables $x_l = \sin^2 \phi_l$ is the ground state wave function for
a quantum many body system on an interval of Calogero-Sutherland type
\cite{BF97a}.

\subsection{Significance of the distribution of the largest or smallest eigenvalue}
Denote by  $E_N^{\rm J}(0;(s,1);\lambda_1,\lambda_2,\beta)$  the probability that there are no
eigenvalues in the interval $(s,1)$ of the Jacobi ensemble (\ref{1}). The density function
$p_{\rm max}^{(N)}(s;\lambda_1,\lambda_2,\beta)$ for the  distribution of the largest eigenvalue
is obtained from $E_N^{\rm J}$ according to
\begin{equation}\label{1.6a}
p_{\rm max}^{(N)}(s;\lambda_1,\lambda_2,\beta) = {d \over ds} E_N^{\rm J}(0;(s,1);\lambda_1,\lambda_2,\beta).
\end{equation}
Since the Jacobi ensemble (\ref{1}) is unchanged by the mappings $x_l \mapsto 1 - x_l$,
$\lambda_1 \leftrightarrow \lambda_2$ we see that
$$
E_N^{\rm J}(0;(s,1);\lambda_1,\lambda_2,\beta) = E_N^{\rm J}(0;(0,s);\lambda_2,\lambda_1,\beta),
$$
and so the density function $p_{\rm min}^{(N)}(s;\lambda_1,\lambda_2,\beta)$ for the distribution of the smallest eigenvalue 
is related to $p_{\rm max}^{(N)}$ by
$$
p_{\rm min}^{(N)}(s;\lambda_1,\lambda_2,\beta) = p_{\rm max}^{(N)}(s;\lambda_2,\lambda_1,\beta).
$$

Interest in $p_{\rm max}^{(N)}$, for the choices of $(\lambda_1,\lambda_2,\beta)$ as relevant to multivariate statistics,
comes by way of Roy's largest root test \cite{An58,Mu82}. Taken literally, this applies to the combination of Wishart
matrices $W_1^{-1} W_2$, where $W_1$ and $W_2$ have the interpretation of covariance matrices within and between
classes respectively; see \cite[\S 2.2]{JN17} for a clear discussion. For a recent work on the
computation of $p_{\rm max}^{(N)}$, specific to the case $\beta=1$ as is of primary interest
in multivariate statistics, see \cite{Ch16}.

The interpretation of the Jacobi ensemble in terms of singular values of a sub-block of a Haar distributed
unitary matrix gives motivation for knowledge of $p_{\rm min}^{(N)}$ \cite{Du12}. Thus particular
randomised algorithms in scientific computing make use of these blocks in linear systems, and so an
important quantity is the corresponding condition number, which is controlled by the smallest
eigenvalue.

In the general case, $p_{\rm min}^{(N)}$ and $p_{\rm max}^{(N)}$ represent examples of extreme
value statistics in a strongly correlated system \cite{MPS20}. For large $N$, and with
tuning of the parameters to a soft edge, large deviation deviation principles apply, giving rise to a macroscopic
viewpoint of the tails \cite{Fo12,MS14}.

\subsection{Our study}
Our aim is to provide computable, structured expressions for
\begin{equation}\label{1.3a}
E_N^{\rm J}(0;(s,1);\lambda_1,\lambda_2,\beta) =
 \int_0^s dx_1 \cdots  \int_0^s dx_N \, {\mathcal P}^{\rm J}(x_1,\dots,x_N).
\end{equation}
This will be done for the parameter ranges
\begin{enumerate}
\item $\lambda_1 > -1, \: \lambda_2 \in \mathbb Z_{\ge 0}, \: \beta > 0$; \nonumber \\
\item $\lambda_1 > -1, \: \lambda_2 = - \beta/2 +k > -1 \: (k \in \mathbb Z_{\ge 0}),\: \beta > 0$; \nonumber \\
\item$ \lambda_1 \in  \mathbb Z_{\ge 0}, \: \lambda_2 > -1,\: \beta \in \mathbb Z_{>0}$.
\end{enumerate}

The starting point for cases (1) and (2) is to change variables $x_l \mapsto s x_l$ in (\ref{1.3a}) and so obtain
\begin{multline}\label{1.3b}
E_N^{\rm J}(0;(s,1);\lambda_1,\lambda_2,\beta) \\ =
{s^{N(\lambda_1 + 1) + \beta N (N-1)/2} \over J_{N,\lambda_1,\lambda_2,\beta}} \int_0^1 dx_1 \cdots  \int_0^1 dx_N \, \prod_{l=1}^N x_l^{\lambda_1} (1 - s x_l)^{\lambda_2}
\prod_{1 \le j < k \le N} | x_k - x_j|^\beta.
\end{multline}
In case (1), in which $ \lambda_2 \in \mathbb Z_{\ge 0}$, we see that the multiple integral in (\ref{1.3b}) is a polynomial of degree $\lambda_2 N$ and thus
\begin{equation}\label{1.3c}
E_N^{\rm J}(0;(s,1);\lambda_1,\lambda_2,\beta) =
s^{N(\lambda_1 + 1) + \beta N (N-1)/2} \sum_{p=0}^{\lambda_2 N} \gamma_p s^p,
\end{equation}
for certain coefficients $\{ \gamma_p \}$. We will show that a known linear differential-difference equation \cite{Fo93} 
from the broader theory of the Selberg integral (see e.g.~\cite[Ch.~4]{Fo10}) can be used to compute these coefficients.

With $\lambda_2 = - \beta/2$ and thus in case (2), although the multiple integral in (\ref{1.3b}) is no longer a polynomial,
it is known to equal a particular classical Gauss hypergeometric function \cite{Ka93, Du12}, \cite[Prop.~13.1.3]{Fo10}
\begin{multline}\label{1.3d}
{1 \over  J_{N,\lambda_1,0,\beta}} \int_0^1 dx_1 \cdots  \int_0^1 dx_N \, \prod_{l=1}^N x_l^{\lambda_1} (1 - s x_l)^{-\beta/2}
\prod_{1 \le j < k \le N} | x_k - x_j|^\beta  \\ =
{}_2 F_1\Big (\beta N/2, (\beta/2)(N-1) + \lambda_1 + 1, \beta (N-1) + \lambda_1 + 2; s \Big ).
\end{multline}
Using this as a seed in the differential-difference equation of \cite{Fo93} leads to the structured expression
\begin{equation}\label{1.3e}
E_N^{\rm J}(0;(s,1);\lambda_1,-\beta/2 + k ,\beta) =
s^{N(\lambda_1 + 1) + \beta N (N-1)/2} \Big ( P(s) f(s) + Q(s) f'(s) \Big ),
\end{equation}
where $f(s)$ is the Gauss ${}_2 F_1$ function in (\ref{1.3d}), and $P(s), Q(s)$ are polynomials of degree less than or
equal to $k N, k(N + 1) $. We remark that an analogous structure, applying to the probability density of the smallest
eigenvalue density in the Laguerre orthogonal ensemble, for Laguerre weight $ \chi_{x > 0} x^{m - 1/2} e^{-x}$
with $m \in \mathbb Z_{\ge 0}$, has been given by Edelman \cite{Ed91}.

We now turn our attention to the parameter range specifying (3). We begin by using the symmetry of the
integrand in (\ref{1.3a})  to order the integration variables
$$
s > x_1 > x_2 > \cdots > x_{N-1} > x_N > 0.
$$
Then 
\begin{equation}\label{1.12}
\prod_{1 \le j < k \le N} | x_k - x_j|^\beta = \prod_{1 \le j < k \le N} ( x_j - x_k)^\beta  =\sum_{\kappa} \alpha_\kappa \prod_{l=1}^N x_l^{\kappa_l},
\end{equation}
for some coefficients $\alpha_\kappa$,
where $\kappa$ is a partition $\kappa_1 \ge \kappa_2 \ge \cdots \ge \kappa_N \ge 0$ involving up to $N$ parts  of fixed length  $\sum_{j=1}^N  \kappa_i = \beta N (N-1)/2$, and further
constrained so that  $\kappa_j \le (N - j)\beta$.

Substituting in (\ref{1.3a}) shows that for $\beta \in \mathbb Z_{\ge 0}$
\begin{equation}\label{10.1}
E_N^{\rm J}(0;(s,1);\lambda_1,\lambda_2,\beta) =
{N! \over J_{N,\lambda_1,\lambda_2,\beta}} 
\sum_{\kappa} \alpha_\kappa
\int_0^s dx_1 \int_0^{x_1} dx_2 \cdots  \int_0^{x_{N-1}} dx_N \, \prod_{l=1}^N x_l^{\lambda_1+ \kappa_l} (1 - x_l)^{\lambda_2}.
\end{equation}
In general,
$$
\int_0^x dX \, X^a(1-X)^b =  \int_0^1 d X \, X^a (1 - X)^b - \int_0^{1-x} d X \, X^b (1 - X)^a.
$$
For $a$ a non-negative integer, the final factor in the second integral can be expanded according to the
binomial theorem, telling us that in this circumstance
\begin{equation}\label{10.1a}
\int_0^x dX \, X^a(1-X)^b = B(a,b)  - \sum_{p=0}^a (-1)^p C_p^a {1 \over b + p + 1} (1-x)^{b+p+1},
\end{equation}
where $C_p^a$ denotes the binomial coefficient, and
\begin{equation}\label{10.1ax}
B(a,b) := \int_0^1 d X \, X^a (1 - X)^b = {\Gamma(a+1) \Gamma (b+1) \over \Gamma(a+b+2)}
\end{equation}
is the Euler beta function.

Making repeated use of (\ref{10.1a}) in (\ref{10.1}) shows that for the parameter range of (3)
\begin{equation}\label{10.1b}
E_N^{\rm J}(0;(s,1);\lambda_1,\lambda_2,\beta) = 1 + \sum_{q=1}^N (1 - s)^{q(\lambda_2 + 1) } \sum_{l=l_{\rm min}(q)}^{l_{\rm max}(q)}
 \gamma_{q,l}  (1 - s)^{l}
\end{equation}
for some $\{ \gamma_{q,l} \}$,
and where $l_{\rm min}(q)$, $l_{\rm max}(q)$ 
are shown below to be given by
\begin{equation}\label{10.1c}
l_{\rm min}(q) =  q(q-1)\beta/2, \qquad
l_{\rm max}(q) = q\lambda_1 + q(N-q)\beta + q(q-1) \beta/2,
\end{equation}
allowing (\ref{10.1b}) to be rewritten
\begin{equation}\label{10.1bx}
E_N^{\rm J}(0;(s,1);\lambda_1,\lambda_2,\beta) = 1 + \sum_{q=1}^N (1 - s)^{q(\lambda_2 + 1)  + q( q -1 ) \beta/2} \sum_{l=0}^{q\lambda_1 + q(N-q)\beta}
 \tilde{\gamma}_{q,l}  (1 - s)^{l}.
\end{equation}
We will show how the differential-difference equation of \cite{Fo93} 
can be adapted to compute $\{  \tilde{\gamma}_{q,k} \}$.
In fact two methods are given, with the second involving the recursive computation of
each of $\{ E_n^{\rm J}(0;(s,1),\lambda_1,\lambda_2,\beta) \}_{n=1}^N$, given $E_1^{\rm J}$ as
the initial condition. We will show in Appendix A how this latter method can be modified to
provide an analogous recursive computational scheme for the gap probability
$E_N^{\rm C}(0;(0,\phi);\beta)$, the probability of no eigenvalues in the $\beta$-circular ensemble,
specified by the class of eigenvalue probability density functions
\begin{equation}\label{10.1d}
{1 \over \mathcal N_{\beta,N}} \prod_{1 \le j < k \le N} | e^{i \theta_k} - e^{i \theta_j} |^\beta.
\end{equation}
Here $0 \le \theta_j \le 2 \pi$, $(j=1,\dots, N)$ and $\mathcal N_{\beta,N}$ is the normalisation; see
e.g.~\cite[Prop.~4.7.2]{Fo10} for its explicit evaluation.

\section{Computation of the coefficients}

\subsection{The differential-difference system}
Our ability to compute the coefficients in the structural forms
(\ref{1.3c}) , (\ref{1.3e}) and (\ref{10.1b}) rests with a recursion scheme satisfied by a
generalisation of the multi-dimensional integral in (\ref{1.3b}).

Let $e_p(y_1,\dots,y_N) $ denote the elementary symmetric polynomials in $\{ y_j \}_{j=1}^N$, and
define
\begin{multline}\label{5.1}
J_{p,N,R}^{(\alpha)}(x) = {1 \over C_p^N}\int_R dt_1 \cdots  \int_R dt_N \,
\prod_{l=1}^N t_l^{\lambda_1} (1 - t_l)^{\lambda_2} (x - t_l)^\alpha \\
\times \prod_{1 \le j < k \le N} | t_k - t_j|^\beta 
e_p(x - t_1,\dots, x - t_N),
\end{multline}
where $R=[0,1]$ or $R=[x,1]$.
 This family of multiple integrals satisfies the
differential-difference system \cite{Fo93}, \cite[\S 4.6.4]{Fo10}\footnote{
These references have $\alpha$ replaced by $\alpha - 1$ relative to our (\ref{5.1}).},
later observed to be equivalent to a certain Fuchsian matrix differential equation \cite{FR12},
\begin{multline}\label{5.1a}
(N - p) E_p J_{p+1}(x) 
= (A_p x + B_p) J_p(x) - x(x-1) {d \over dx} J_p(x) + D_p x ( x - 1) J_{p-1}(x),
\end{multline}
where we have abbreviated $J_{p,N,R}^{(\alpha)}(x) =: J_p(x)$, and
\begin{align*}
A_p & = (N-p) \Big ( \lambda_1 + \lambda_2 + \beta (N - p - 1) + 2(\alpha + 1) \Big ) \\
B_p & = (p-N)  \Big ( \lambda_1 + \alpha + 1 + (\beta/2) (N - p - 1) \Big ) \\
D_p & = p \Big ( (\beta/2)(N-p) + \alpha + 1 \Big ) \\
E_p & = \lambda_1 + \lambda_2 + 1 + (\beta/2) (2N - p - 2) + (\alpha + 1).
\end{align*}

Starting with knowledge of $J_{p,N,R}^{(\alpha)}(x) |_{p=0}$, application of the recurrence
for $p=0,1,\dots,N$ gives us an expression for $J_{0,N,R}^{(\alpha+1)}(x)$, since
from the definition (\ref{5.1})
\begin{equation}\label{5.1b}
J_{p,N,R}^{(\alpha)}(x) \Big |_{p=N} = J_{0,N,R}^{(\alpha+1)}(x).
\end{equation}
Repeating this, we can deduce an expression for $J_{0,N,R}^{(\alpha+k)}(x)$ from
 knowledge of $J_{0,N,R}^{(\alpha)}(x)$ for any positive integer $k$.
 
 \subsection{Cases (1) and (2)}
 
For application to the parameter ranges (1) and (2) as listed below (\ref{1.3a})
we set $R=[0,1]$ in (\ref{5.1}) and consider the transformed integrals
\begin{multline}\label{6.1}
\tilde{J}_{p,N}^{(\alpha)}(x) = {x^{N \alpha + p} \over J_{N,\lambda_1,\lambda_2,\beta}} J_{p,N,R}^{(\alpha)}(1/x) \Big |_{R=[0,1]} \\
=
{1 \over J_{N,\lambda_1,\lambda_2,\beta} C_p^N}
 \int_0^1 dt_1 \cdots  \int_0^1 dt_N \,
\prod_{l=1}^N t_l^{\lambda_1} (1 - t_l)^{\lambda_2} (1 - x t_l)^\alpha \\
\times \prod_{1 \le j < k \le N} | t_k - t_j|^\beta 
e_p(1 - xt_1,\dots, 1 - xt_N).
\end{multline}
The normalisations are such
\begin{equation}\label{6.1s}
\tilde{J}_{p,N}^{(\alpha)}(x) \Big |_{x=0} = \tilde{J}_{p,N}^{(\alpha)}(x)  \Big |_{p=\alpha =0}=1,
\end{equation}
and we have, using (\ref{1.3b}),
\begin{equation}\label{6.1t}
E_N^{\rm J}(0;(s,1);\lambda_1,\lambda_2,\beta)  
=s^{N(\lambda_1 + 1) + \beta N (N-1)/2}
{J_{N,\lambda_1,0,\beta} \over J_{N,\lambda_1,\lambda_2,\beta}} \cdot \tilde{J}_{0,N}^{(\alpha)}(s) \Big |_{\lambda_2 = 0, \alpha \mapsto \lambda_2 }.
\end{equation}
In addition to the closed form evaluation (\ref{1.3d}) as is
relevant to case (2), we also have that \cite{Ao87},  \cite[Prop.~13.1.2]{Fo10}
\begin{multline}\label{5.1c}
\tilde{J}_{p,N}^{(\alpha)}(x) \Big |_{p=0, \lambda_2=0,\alpha = 1} 
\\ =
{}_2 F_1\Big (-N, -(N-1) - (2/\beta) (\lambda_1 + 1) , -2(N-1) - (2/\beta) (\lambda_1 +2); x \Big ),
\end{multline}
which is relevant to case (1).

In terms of the  transformed integrals (\ref{6.1}), and with the
further abbreviation of notation $\tilde{J}_{p,N}^{(\alpha)}(x) =:  \tilde{J}_{p}(x)$
analogous to what was used in (\ref{5.1a}),
 the differential-difference system (\ref{5.1a}) reads
\begin{multline}\label{7.1a}
(N - p) E_p \tilde{J}_{p+1}(x) \\
= (A_p - (N \alpha + p)  + (B_p + N \alpha + p)x) \tilde{J}_p(x) + x(1-x) {d \over dx} \tilde{J}_p(x) + D_p  ( 1 - x) \tilde{J}_{p-1}(x).
\end{multline}
In case (1), the recurrence is begun with $\lambda_2 = 0$, $\alpha = 0$ using the second of the equations in (\ref{6.1s}) which is the
$\lambda_2 = 0$ parameter value in (1). After iterating for
$p=0,1,\dots,N$, use is made of the analogue of (\ref{5.1b}) to now have at hand
the form of the polynomial $\tilde{J}_{0,N}^{(\alpha)}(x) |_{\lambda_2 = 0,\alpha = 1}$ (which must
agree with (\ref{5.1c})). This procedure is then repeated a total of $\lambda_2$ times to
eventually compute $\tilde{J}_{0,N}^{(\alpha)}(x) |_{\lambda_2 = 0,\alpha \mapsto \lambda_2}$.
After multiplication by a suitable normalisation, as made explicit in (\ref{6.1t}), we have available
all the coefficients $\{ \gamma_p \}$ in (\ref{1.3c}). Note that since $E_N^{\rm J}(0;(s,1);\lambda_1,\lambda_2,\beta) |_{s=1} = 1$,
we must have the sum rule
\begin{equation}\label{s1}
\sum_{p=0}^{N \lambda_2} \gamma_p = 1,
\end{equation}
 which provides a useful check on the
computation. Another check follows from the fact that in addition to (\ref{5.1a}), the
family of integrals (\ref{5.1}) in the polynomial case $R=[0,1]$, $\alpha \in \mathbb Z_{\ge 0}$
also satisfy a multidimensional difference equation \cite[Eq.~(30)]{FI10a}, which allows
for an independent computation.

We now turn our attention to case (2). First, we will show how (\ref{7.1a}), initialised with $\alpha = -\beta/2$ so that
$ \tilde{J}_{0}$ is given by (\ref{1.3d}), implies the structure (\ref{1.3e}). An essential point is that 
generally the Gauss hypergeometric function ${}_2 F_1(a,b,c;x)$ satisfies the second order linear
differential equation
\begin{equation}\label{7.1b}
x (1 - x) {d^2 y \over dx^2} + \Big ( c - (a + b + 1) x \Big ) {d y \over dx} - a b y = 0.
\end{equation}
Denote the Gauss hypergeometric function in (\ref{1.3d}) by $f(a,b,c;s) =: f(s)$ as is consistent with the
notation used in (\ref{1.3e}). For some functions $P_p$, $Q_p$ write
\begin{equation}\label{7.1c}
\tilde{J}_p(x) = P_p(x) f(x) + Q_p(x) f'(x).
\end{equation}
Substituting in (\ref{7.1a}) and making use of (\ref{7.1b}) shows
\begin{multline}\label{8.1a}
(N-p) E_p P_{p+1}(x) = \Big ( A_p - (N \alpha + p) + (B_p + N \alpha + p) x \Big ) P_p(x) \\
+ D_p (1 - x) P_{p-1}(x) +
x(1 - x) P_p'(x) + a b Q_p(x) 
\end{multline}
and
\begin{multline}\label{8.1b}
(N-p) E_p Q_{p+1}(x) \\
= \Big ( A_p - (N \alpha + p) + (B_p + N \alpha + p) x \Big ) Q_p(x) + D_p (1 - x) Q_{p-1}(x)\\
+  x (1 - x) ( P_{p}(x) + Q_p'(x)) - ( c - (a + b + 1) x) Q_p(x).
\end{multline}
Now  (\ref{1.3d}) and (\ref{6.1}) tells us that for $\alpha = -\beta/2$, $\lambda_2 = 0$ we have $(P_0(x), Q_0(x)) = (1,0)$, so we see by
iteration of (\ref{8.1a}) and (\ref{8.1b}) that (\ref{1.3e}) holds true, with $P(s), Q(s)$ polynomials
of the stated degree. The recursion is initialised with $\tilde{J}_0^{(\alpha)}|_{\lambda_2=0,\alpha=-\beta/2}$ and then (\ref{8.1a}) and (\ref{8.1b}) are iterated for $p=0,1,...,N$ to obtain $\tilde{J}_N^{(\alpha)}|_{\lambda_2=0,,\alpha=-\beta/2}$. The latter is identical to $\tilde{J}_0^{(\alpha)}|_{\lambda_2=0,\alpha=-\beta/2+1}$, thereby resulting in the increase of $\alpha$ by 1. This procedure is repeated $k$ times to eventually obtain $\tilde{J}_N^{(\alpha)}|_{\lambda_2=0,\alpha=-\beta/2+k}$, which in turn results in the desired gap probability expression using (\ref{6.1t}).

A number of checks on the computation are possible. One is that
\begin{equation}\label{s2}
\lim_{s \to 1^-}   \Big ( P(s) f(s) + Q(s) f'(s) \Big ) =1,
\end{equation}
as analogous to (\ref{s1}). Moreover, for $\beta$ even, and
$k$ a positive integer such that $-\beta / 2 + k \ge 0$, we see that case
(2) coincides with case (1), and in particular the structure (\ref{1.3c}) must hold true
with $\lambda_2 = -\beta / 2 + k$.

\begin{figure}[!ht]
\centering
\includegraphics[width=3in]{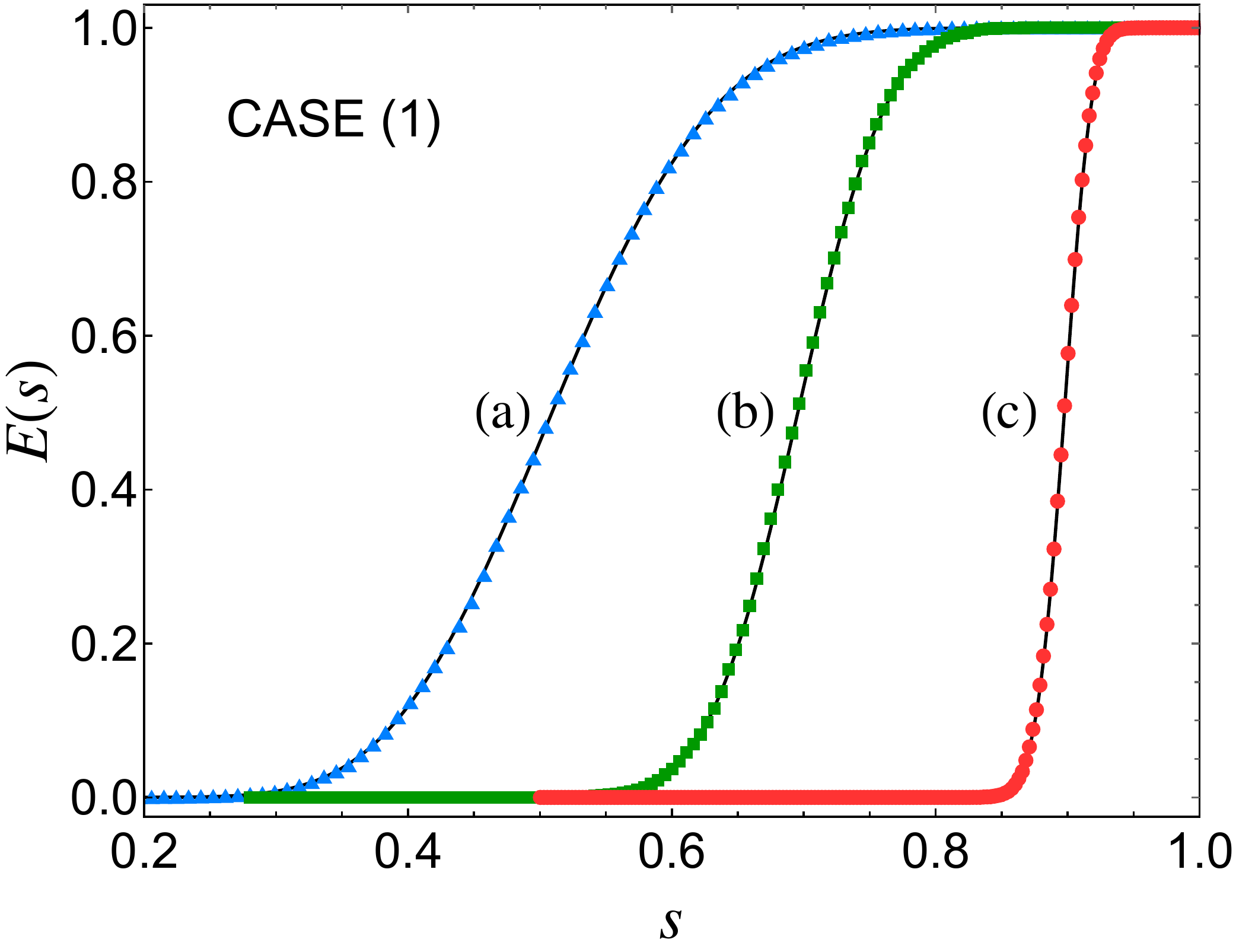}
\caption{Gap probabilities $E_N^{\rm J}(0;(s,1);\lambda_1,\lambda_2,\beta)\equiv E(s)$ in case (1) for three sets of parameter values: (a) $n = 7, \lambda_1 = -3/4, \lambda_2 = 9, \beta = 7/8$; (b) $n = 10, \lambda_1 = 2, \lambda_2 = 15, \beta = 3/2$; (c) $n = 15, \lambda_1 = 5, \lambda_2 = 25, \beta = 5$. The solid curves have been obtained using the recursive scheme and symbols are based on Monte Carlo simulation of matrix model (\ref{5.1x}).}
\label{figcase1}
\end{figure}

\begin{figure}[!h]
\centering
\includegraphics[width=3in]{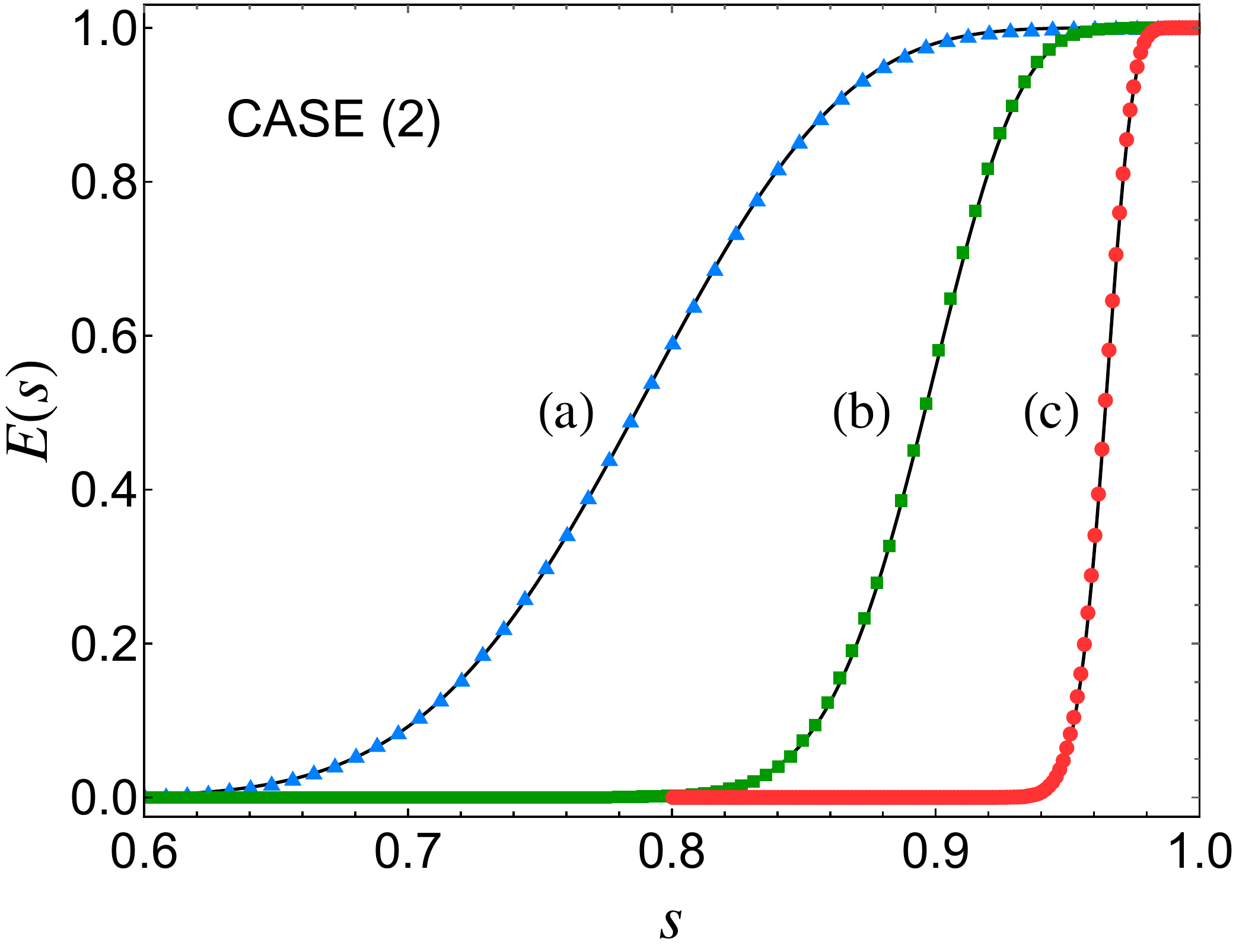}
\caption{Gap probabilities $E_N^{\rm J}(0;(s,1);\lambda_1,\lambda_2,\beta)\equiv E(s)$ in case (2) for three sets of parameter values: (a) $n = 5, \beta = 1/2, \lambda_1 = 9, k = 6, \lambda_2 = -\beta/2 + k=23/4$; (b) $n = 9, \beta = 8/3, \lambda_1 = 7, k = 10, \lambda_2 = -\beta/
   2 + k=26/3$; (c) $n = 18, \beta = 5, \lambda_1 = 16/3, k = 16, \lambda_2 = -\beta/2 + k=27/2$.}
   \label{figcase2}
\end{figure}

In figures \ref{figcase1} and \ref{figcase2}, we show gap probabilities for the cases (1) and (2) obtained using the recursive schemes described above. We consider three sets of parameters for each of these cases. Moreover, for comparison, we also show the results obtained using the Monte Carlo simulations of the matrix model (\ref{5.1x}) as overlaid symbols and agree perfectly with the analytical results depicted using solid curves. A Mathematica~\cite{Mtmk} file with codes implementing the above recursive schemes is provided as an ancillary file.


\subsection{Case (3)}\label{S2.3}

We now deal with the parameter range (3) as listed below (\ref{1.3a}) and, to begin with, we focus on even $\beta$. Eventually, we will discuss a strategy to handle a general $\beta\in {\mathbb Z_{>0}}$. For the present case, we require the family of integrals
(\ref{5.1}) with both $R=[0,1]$ and $R=[x,1]$:
\begin{equation}\label{s2a}
\int_R d X \mapsto \Big ( \int_0^1 + \zeta  \int_x^1 \Big ) d X.
\end{equation}
This choice of $R = R(\zeta)$ relates to $E_N^{\rm J}(0;(s,1);\lambda_1,\lambda_2,\beta)$ according to
\begin{equation}\label{s3z}
E_N^{\rm J}(0;(s,1);\lambda_1,\lambda_2,\beta) = {1 \over J_{N,\lambda_1,\lambda_2,\beta} }
J_{0,N,R(\zeta)}^{(\alpha)}(x)\Big |_{\alpha = 0, \zeta = -1}.
\end{equation}

The relevance of introducing this seemingly complicated choice of $R$ comes from
a viewpoint of the differential-difference system (\ref{5.1a}) as a matrix linear
differential equation \cite{FR12}. Thus with $R(\zeta)$ as 
specified by (\ref{s2a}), we read off from (\ref{5.1a}) that the vector of integrals
\begin{equation}\label{s4}
{\mathbf J}_{N}^{(\alpha)}(x;\zeta) = {1 \over J_{N,\lambda_1,\lambda_2,\beta} }
[ J_{p,N,R(\zeta)}^{(\alpha)}(x)]_{p=0}^{N}
\end{equation}
satisfies the matrix differential equation
\begin{equation}\label{s3a}
{d \over dx} {\mathbf J} = \bigg ( {\mathbf Z_{-1} \over x - 1} + {\mathbf Z_0 \over x}
+ \mathbf Y \bigg ) \mathbf J,
\end{equation}
where $\mathbf Z_{-1}, \mathbf Z_0, \mathbf Y$ are $(N+1) \times (N+1)$ matrices specified by
\begin{align*}
\mathbf Z_{-1} & = {\rm diag} \, [A_p + B_p]_{p=0}^N - {\rm diag}^+ \, [(N-p) E_p]_{p=0}^{N-1} \\
\mathbf Z_{0} & = {\rm diag} \, [- B_p]_{p=0}^N + {\rm diag}^+ \, [(N-p) E_p]_{p=0}^{N-1} \\
\mathbf Y & = {\rm diag}^- \, [D_p]_{p=1}^N.
\end{align*}
Here ${\rm diag}^+$ refers to the first diagonal above the main diagonal, and
${\rm diag}^-$ the first diagonal below; all other entries are zero.

The matrix differential equation (\ref{s3a}) admits Frobenius type solutions
\begin{equation}\label{s3b}
[ \mathbf w_p^{(q)}(x)]_{p=0}^N, \qquad  \mathbf w_p^{(q)}(x) := (1-x)^{\mu_q} 
\Big [ \sum_{l=0}^\infty c_{l,p}^{(q)} (1-x)^l \Big ]_{p=0}^N,
\end{equation}
where $\mu_q = (A_q + B_q) = (N-q)\Big (\lambda_2 + (\beta/2)(N - q - 1) + 1 \Big )$.
Let us write $\mathbf c_l^{(q)} = [ c_{l,p}^{(q)} ]_{p=0}^{N}$. Substituting 
(\ref{s3b}) in (\ref{s3a}), expanding both sides in a power series in $(1-x)$, and
equating like coefficients shows
\begin{equation}\label{s3d}
\Big (  \mathbf Z_{-1}  - ( \mu_q + n ) \mathbb I_{N+1}  \Big ) \mathbf c_n^{(q)}  =
(\mathbf Z_0 + \mathbf Y) \mathbf c_{n-1}^{(q)}
+ \mathbf Z_0
\Big ( \sum_{s=1}^{n-1}   \mathbf c_{n-1-s}^{(q)} \Big ),
\end{equation}
valid for $n \ge 1$. For $n=0$ the right hand side is zero, telling us that 
$  \mathbf c_{0}^{(q)}$ is the eigenvector of $ \mathbf Z_{-1}$ with eigenvalue
$A_q + B_q$. Only the first $q$ components are non-zero, and satisfy the recurrence
\begin{equation}\label{s3e}
 c_{0,p+1}^{(q)} = {A_p + B_p - A_q - B_q \over (N - p ) E_p}  c_{0,p}^{(q)}, \quad (p=0,\dots, q-1),
 \end{equation}
 where we are free to choose
 \begin{equation}\label{s3f}
  c_{0,0}^{(q)} = 1.
  \end{equation} 
With this as the initial condition, we can use (\ref{s3d}) to
recursively compute the coefficients in (\ref{s3b}), provided
$  \mathbf Z_{-1}  - ( \mu_q + n ) \mathbb I_{N+1} $ is invertible for
$n \ge 1$; for this latter requirement it is sufficient that
$\lambda_2$ be a generic parameter. Moreover, for the $\lambda_2$ values which result in a singular $\mathbf Z_{-1}  - ( \mu_q + n ) \mathbb I_{N+1}$, the recursion scheme may be implemented with $\lambda_2$ kept as a variable. The correct result can then be produced from the final expression by assigning $\lambda_2$ the desired value. However, this does slow down the recursive scheme considerably, if the parameter values are large. Therefore, in such situations, it is generally better to implement an alternative recursion scheme, as discussed ahead.

With $[\zeta^{q}]f$ denoting the coefficient of $\zeta^{q}$ in $f$,
from the definitions (\ref{5.1}) and (\ref{s2a})
\begin{multline}\label{11.1}
[\zeta^{q}]  J_{p,N,R(\zeta)}^{(\alpha)}(x) \Big |_{\alpha = 0} =
{C^N_q \over C^N_p} \int_x^1 dt_1 \cdots  \int_x^1 dt_q 
\int_0^1 dt_{q+1} \cdots  \int_0^1 dt_N  \\
\times \prod_{l=1}^N t_l^{\lambda_1} (1 - t_l)^{\lambda_2}
\prod_{1 \le j < k \le N} | t_k - t_j|^\beta e_p(x-t_1,\dots, x - t_N).
\end{multline}
According to (\ref{s3z}),
 \begin{equation}\label{s4a}
 E_N^{\rm J}(0;(s,1);\lambda_1,\lambda_2,\beta) =
 {1 \over J_{N,\lambda_1,\lambda_2,\beta}}
 \sum_{q=0}^N (-1)^{q}
 [\zeta^{q}]  J_{p,N,R(\zeta)}^{(\alpha)}(x) \Big |_{\alpha =p=0 } .
 \end{equation}
We know that independent of $q$, the family of multiple integrals (\ref{11.1}) satisfies
the matrix differential equation (\ref{s3a}). Moreover a simple change of variables
shows that for $\beta$ a non-negative even integer, the power series in $(1-x)$
has the structure (\ref{s3b}) with $q$ replaced by $N-q$, 
and furthermore $c_{l,p}^{(N-q)}|_{p=0} = 0$ for $l > l_{\rm max}(q)$, where
$l_{\rm max}(q)$ is given by (\ref{10.1c}), thus providing justification of the latter, and also
that  $c_{l,p}^{(N-q)}|_{p=0} = 0$ for $l < l_{\rm min}(q)$, where
$$
l_{\rm min}(q) = q (q-1)\beta/2.
$$
In particular, it must be that (\ref{11.1}) is
proportional to (\ref{s3b}) with $q$ replaced by $N-q$.
 If we normalise (\ref{s3b}) so that $c_{0,0}^{(N-q)} = 1$, consideration
of the small $x$ behaviour of the multiple integral corresponding to 
$J_{p,N,R(\zeta)}^{(\alpha)}(x) |_{\alpha = 0, p = 0} $ allows us to compute
the proportionality and so conclude
$$
[\zeta^{q}] {\mathbf J}_{N}^{(\alpha)}(x;\zeta) \Big |_{\alpha = 0} = 
C_q^N {J_{N-q,\lambda_2+q \beta ,\beta} J_{q, \lambda_2, 0,\beta} \over J_{N,\lambda_1,\lambda_2,\beta}}
[ \mathbf w_p^{(N-q)}(x)]_{p=0}^N.
$$
This holds true for general $\lambda_1 > -1$; the significance of restricting
to $\lambda_1$ a non-negative integer is that then the power series
in (\ref{s3b}) terminates. We remark that in the case $\beta = 2$ an
alternative way to compute the expansion (\ref{10.1b}) is to use the
characterisation of $E_N^{\rm J}$ in that case in terms of the
solution of a particular Painlev\'e VI equation in sigma form;
see \cite[\S 8.3.1 and eq.~(8.77)]{Fo10}. Working overlapping with the
above in the special case $\beta = 1$ can also be found in earlier literature. Specifically, a matrix differential equation equivalent to (\ref{s3a}) for the
corresponding family of integrals (\ref{11.1}) was derived by Davis
\cite{Da72} in a pioneering paper on this characterisation published in 1972.

There is an alternative method that can be used for the same purposes of computing
the expansion (\ref{s3b}) and to handle $\beta\in {\mathbb Z_{>0}}$. This is to follow a procedure detailed
in our recent work \cite{FK19} for the $\beta$-Laguerre ensemble, which itself can be
traced back to \cite{Ja75} where it was used to give a recursive computation of the
probability density function for the Gaussian orthogonal ensemble;
for the recursion corresponding to (\ref{5.1a}) for the $\beta$-Laguerre ensemble
see \cite{Ku19}, \cite{FT19}.
For $\lambda_1$ a non-negative integer, and general $\lambda_2 > - 1$,
we begin with (\ref{10.1a}) to compute
$\int_0^s dt_1 \, t_1^{\lambda_1} (1 - t_1)^{\lambda_2}$ in the required form.
Applying the recurrence (\ref{5.1a}) with $\alpha = 0,\dots, \beta - 1$ gives
the evaluation of 
\begin{equation}\label{8.1d}
\int_0^s dt_1 \, t_1^{\lambda_1} (1 - t_1)^{\lambda_2} (s - t_1)^\beta,
\end{equation}
with the structure (\ref{10.1a}) now replaced by $Q(s) + (1 - s)^{\lambda_2}
R(s)$ for polynomials $Q(s)$, $R(s)$ in $(1-s)$.
 In this we set $s = t_2$, multiply by
$(1 - t_2)^{\lambda_2} t_2^{\lambda_1}$, and integrate over $t_2$ from $0$
to $s$ using (\ref{10.1a}) again, thereby evaluating $E_N^{\rm J}(0;(s,1);\lambda_1,\lambda_2,\beta)$ for 
$N=2$  with the structure (\ref{10.1b}). Repeating this procedure, we
see that for the parameter range (3) we can compute
$E_N^{\rm J}(0;(s,1);\lambda_1,\lambda_2,\beta)$ for any fixed $N$ from knowledge of
$E_{N-1}^{\rm J}(0;(s,1);\lambda_1,\lambda_2,\beta)$, and moreover the final 
expression will have the structure (\ref{10.1b}).

A sum rule associated with (\ref{10.1b}) is to use the fact that
$E_N^{\rm J}(0;(s,1);\lambda_1,\lambda_2,\beta) |_{s=0} = 0$ to conclude
\begin{equation}\label{s7}
0 = 1 + \sum_{q=1}^N \sum_{l=0}^{q \lambda_1 + q (N - q)\beta}\tilde{ \gamma}_{q,l}.
\end{equation}
Also, for $\lambda_2 = - \beta/2 + k > -1$, agreement must be obtained
with case (3), and for $\lambda_2 \in \mathbb Z_{\ge 0}$ when (\ref{8.1a}) is 
a polynomial in $s$, the result must agree with that obtained in case (1).

\begin{figure}[!h]
\centering
\includegraphics[width=3.5in]{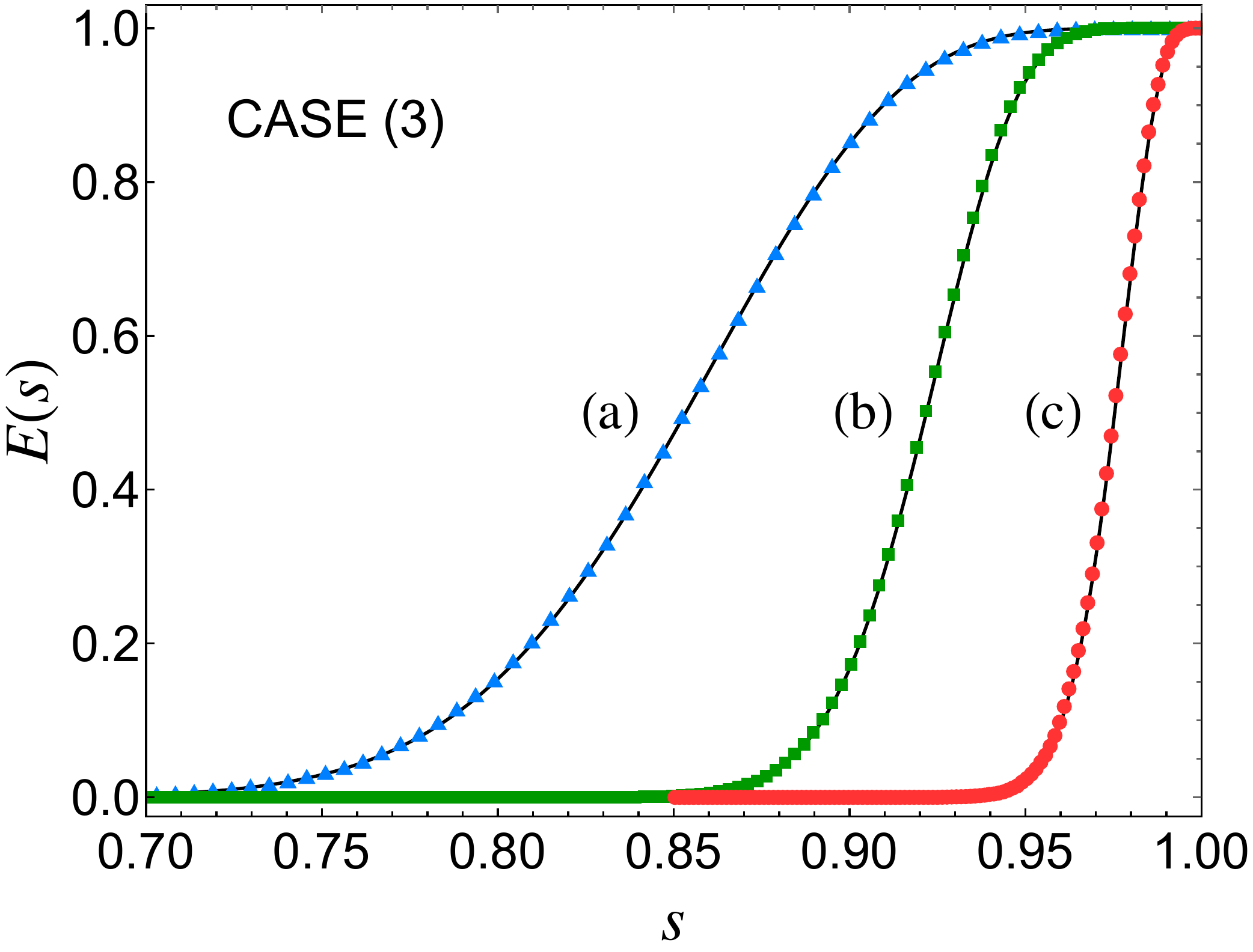}
\caption{Gap probabilities $E_N^{\rm J}(0;(s,1);\lambda_1,\lambda_2,\beta)\equiv E(s)$ in case (3) for three sets of parameter values:  (a) $n = 6, \lambda_1 = 5, \lambda_2 = 17/3, \beta = 2$; (b) $n = 11, \lambda_1 = 1, \lambda_2 = 49/5, \beta = 4$; (c) $n = 25, \lambda_1 = 4, \lambda_2 = 32/9, \beta = 1$.}
\label{figcase3}
\end{figure}

In figure \ref{figcase3}, we show the gap probabilities obtained using the above described recursive approaches in the case (3) for three sets of parameter values. The first two sets can be handled by the recursion scheme involving matrices given by (\ref{s3d}), whereas for the last one, we require the alternative recursive approach. As for figures \ref{figcase1} and \ref{figcase2}, Monte Carlo simulation results based on (\ref{5.1x}), shown as overlaid symbols, are found to be in perfect agreement with the solid curves obtained using the recursive schemes. Mathematica~\cite{Mtmk} files with codes implementing the above described recursion schemes are attached as ancillary files.

\subsection{Direct computation of $p_{\rm max}^{(N)}$}

Our concern in the above sections has been with the computation of 
$E_N^{\rm J}(0,(s,1);\lambda_1,\lambda_2,\beta)$. The probability density
function $p_{\rm max}^{(N)}$ for the distribution of the largest eigenvalue
is then computed by differentiation according to (\ref{1.6a}). Here we consider
the task of computing $p_{\rm max}^{(N)}$ directly for the
ranges of parameters (1) and (2) as listed below (\ref{1.3a}).

Differentiating (\ref{1.3a}) gives
$$
p_{\rm max}^{(N)}(s,\lambda_1,\lambda_2,\beta) = N \int_0^s dx_2 \cdots  \int_0^s dx_N \, \mathcal P^{\rm J}(s,x_2,\dots,x_N).
$$
Changing variables $x_l \mapsto s x_l$ shows that analogous to (\ref{1.3b})
\begin{multline}\label{1.3bx}
p_{\rm max}^{(N)}(s;\lambda_1,\lambda_2,\beta)  ={N s^{N(\lambda_1 + 1) + \beta N (N-1)/2-1} (1 - s )^{\lambda_2}  \over J_{N,\lambda_1,\lambda_2,\beta}} \\
\times
\int_0^1 dx_2 \cdots  \int_0^1 dx_N \, \prod_{l=2}^N x_l^{\lambda_1} (1 - x_l)^\beta (1 - s x_l)^{\lambda_2}
\prod_{2 \le j < k \le N} | x_k - x_j|^\beta.
\end{multline}
It follows that for $\lambda_2 \in \mathbb Z_{\ge 0}$, which is case (1) of the parameter ranges,  we have the structure
\begin{equation}\label{1.10x}
p_{\rm max}^{(N)}(s;\lambda_1,\lambda_2,\beta)  = s^{N(\lambda_1 + 1) + \beta N (N-1)/2-1} (1 - s )^{\lambda_2} \sum_{p=0}^{(\lambda_2 - 1)N} \tilde{\gamma}_p s^p.
\end{equation}
Note that this analytic form is consistent with differentiating (\ref{1.3c}) with respect to $s$.
By appropriate choice of the parameters the recurrence (\ref{7.1a}) can be applied to directly compute
$\{ \tilde{\gamma}_p \}$ in (\ref{1.10x}).

For the parameter range (2), we require a modification of (\ref{1.3d}),
\begin{multline}\label{1.3dx}
{1 \over  J_{N-1,\lambda_1,\beta,\beta}} \int_0^1 dx_2 \cdots  \int_0^1 dx_N \, \prod_{l=2}^N x_l^{\lambda_1} (1 - s x_l)^{-\beta/2}
\prod_{2 \le j < k \le N} | x_k - x_j|^\beta  \\ =
{}_2 F_1\Big (\beta (N - 1)/2, (\beta/2)(N-2) + \lambda_1 + 1, \beta (N-1) + \lambda_1 + 2; s \Big );
\end{multline}
see e.g.~\cite[Prop.~13.1.3]{Fo10}. Using this as a seed in the recurrence (\ref{7.1a}), the method of
the paragraph including (\ref{7.1b}) allows us to compute polynomials $\tilde{P}(s), \tilde{Q}(s)$ such
that
\begin{equation}\label{9.1x}
p_{\rm max}^{(N)}(s;\lambda_1,\lambda_2,\beta)  = 
 s^{N(\lambda_1 + 1) + \beta N (N-1)/2-1} (1 - s )^{\lambda_2} 
 \Big ( \tilde{P}(s) \tilde{f}(s) +  \tilde{Q}(s)  \tilde{f}'(s) \Big ),
 \end{equation}
 where $ \tilde{f}(s) $ denotes the Gauss hypergeometric function in (\ref{1.3dx}).
 
 \subsection{The large $N$ limit}
 The distributions $p_{\rm max}^{(N)}$ and $p_{\rm min}^{(N)}$ permit two distinct
 large $N$ limits. One is to a hard edge state, which is dependent on the respective hard
 edge parameter: $\lambda_2$ for the largest eigenvalue, replaced by $\lambda_1$ for
 the smallest eigenvalue. As an explicit example, we see from (\ref{1.3bx}) that
 $$
 \lim_{N \to \infty} {1 \over N^2} p_{\rm max}^{(N)} \Big ( \Big (1 - {s \over N^2} \Big ); \lambda_1, \lambda_2,
 \beta \Big ) \Big |_{\lambda_2 = 0} = {2 \over \beta} e^{- \beta s / 2}.
 $$
 However the dependence on $\lambda_2$ cannot be made explicit in general. Exceptions are for
 $\beta = 1,2$ and 4 when the limiting hard edge distributions can be characterised
 in terms of both Painlev\'e transcendents and Fredholm determinants; see e..g.~\cite[\S 9.5.1, \S 9.8]{Fo10}. 
 The Fredholm form is particularly well suited for numerical computations
 \cite{Bo10}.
 For positive integer and half integer ($\ge -1/2$)
 values of $\lambda_2$ in the case $\beta = 1$ there is a Pfaffian form \cite{AGKWW14},
 and similarly for $\lambda_2$ even in the case $\beta = 4$ \cite{NF98}. The limiting structure
 is even simpler for $\lambda_2$ a positive integer, when it can be
 expressed as a determinant \cite{FH94,MMM19}.
 
 In the circumstance that $\lambda_2$ is proportional to $N$, the limiting support
 of the eigenvalue density is bounded away from 1. An appropriate scaling 
 \cite{Jo08, HF12} gives a well  defined limiting distribution, corresponding to what is termed a 
 soft edge state. This limiting distribution is independent of $\lambda_1, \lambda_2$. For
 $\beta = 1,2$ and 4 it can be characterised in terms of Painlev\'e transcendents 
and Fredholm determinants  \cite{TW94c}. An analogous, but more complicated
characterisation is also known for $\beta = 6$ \cite{Ru16}.

A future application of our recursive, finite $N$, computations of the present paper
is in relation to the theme of scaling this variables so as to obtain an optimal 
rate of convergence to the limiting distributions. In special cases, these scalings
are quantified in \cite{MMM19} for the hard edge limit, and in \cite{Jo08} for the
soft edge limit. In the recent works \cite{FT19a,FT19}, recursions applying for the
$\beta$-Laguerre ensemble have been used to exhibit the optimal rates 
beyond the classical cases $\beta = 1,2$ and 4.

%

   \section*{Acknowledgements}
 P.J.F.~acknowledges support from the Australian Research Council
 (ARC) through the ARC Centre of Excellence for Mathematical \& Statistical Frontiers.

\appendix
\section*{Appendix}
\renewcommand{\thesection}{A} 
\setcounter{equation}{0}
The $\beta$-circular ensemble (\ref{10.1d}) is a particular case of the (generalised)
circular Jacobi $\beta$-ensemble. The latter is specified by the 
family of probability density functions on the unit circle proportional to
\begin{equation}\label{A1}
\prod_{l=1}^N \chi_{-\pi < \theta_l < \pi} e^{b_2 \theta_l}
|1 + e^{i \theta_l} |^{2 b_1} \prod_{1 \le j < k \le N} 
| e^{i \theta_k} - e^{i \theta_j} |^\beta;
\end{equation}
thus we set $b_1 = b_2 = 0$. Introducing $\xi_l = e^{i \theta_l}$ $(l=1,\dots,N)$,
and temporarily requiring that $b_1$ and $\beta/2$ be positive integer, the measure
associated with (\ref{A1}) maps to the measure proportional to
\begin{equation}\label{A2}
\prod_{l=1}^N \xi_l^{\tilde{\lambda}_1} (1 - \xi_l)^{\tilde{\lambda}_2} 
\prod_{1 \le j < k \le N} (\xi_k - \xi_j)^\beta d \xi_1 \cdots d \xi_N
\end{equation}
with
\begin{equation}\label{A2a}
\tilde{\lambda}_1 = - b_1 - 1 - (\beta/2) (N-1), \qquad \tilde{\lambda}_2 = b_1 + i b_2.
\end{equation}
In particular, this tells us that the averages
\begin{equation}\label{A2b}
\Big \langle \prod_{l=1}^N ( x- e^{i \theta_l})^\alpha s_\kappa(e^{i \theta_1},\dots,e^{i \theta_N}) \Big \rangle
\end{equation}
with respect to (\ref{A1}) satisfy the same recurrences (\ref{5.1a}) as the corresponding averages
(\ref{5.1}) for the Jacobi ensemble. We remark that this same conclusion can be reached by
direct application of integration by parts as used in \cite{Fo93}, \cite[\S 4.6]{Fo10}, without the need
to assume  $b_1$ and $\beta/2$ are positive integers.

We would like to make use of the recurrences satisfied by (\ref{A2b}) to provide a recursive computational
scheme for the circular ensemble gap probability
\begin{equation}\label{A2c}
E_N^{\rm C}(0;(0,\phi);\beta) = \lim_{\mu \to 0}
{1 \over \mathcal N_{\beta,N}} \int_\phi^{2 \pi} d \theta_1 \cdots  \int_\phi^{2 \pi} d \theta_N \,
\prod_{l=1}^N e^{i \theta_l \mu} \prod_{1 \le j < k \le N} | e^{i \theta_k} - e^{i \theta_j} |^\beta,
\end{equation}
where the parameter $\mu$ is introduced for later convenience.
As for the Jacobi gap probability $E_N^{\rm J}(0;(s,1);\lambda_1,\lambda_2,\beta)$ in the parameter
range (3), the case of $\beta$ a positive integer is special in this regard. Thus with the ordering
\begin{equation}\label{A2x}
R_N: \: \: 2 \pi > \theta_1 > \theta_2 > \cdots > \theta_N > \phi,
\end{equation}
analogous to (\ref{1.12}) we have
\begin{equation}\label{A2y}
\prod_{1 \le j < k \le N} | e^{i \theta_k} - e^{i \theta_j} |^\beta =
\chi \prod_{l=1}^N e^{-i\theta_l \beta (N-1)/2}
\prod_{1 \le j < k \le N} (e^{i \theta_j} - e^{i \theta_k} )^\beta, 
\end{equation}
where here $\chi$ is a phase, $|\chi| = 1$, which has a polynomial structure. In particular for
$\beta$ a positive integer, the multidimensional integral is a finite series in powers of
$e^{i \phi}$, although taking the limit $\mu \to 0$ will introduce factors which are polynomials
in $\phi$ itself.

From the working of the above paragraph, it suffices to specify a computational scheme for the
integrals
\begin{equation}\label{A2d}
\int_{R_N} d \theta_1 \cdots d \theta_N \, \prod_{l=1}^N e^{i \theta_l \tilde{\mu}}
 \prod_{1 \le j < k \le N} ( e^{i \theta_j} - e^{i \theta_k} )^\beta.
 \end{equation}
 This is done as for the computation of $E_N^{\rm J}(0;(s,1);\lambda_1,\lambda_2,\beta)$
 in the parameter range (3), as detailed in the discussion about (\ref{8.1d}). Actually it is
 a little simpler, since instead of making repeated use of (\ref{10.1a}), we only require use of
 $$
 \int_0^\phi d \theta \, e^{i \theta \nu}  = {1 \over i \nu } (e^{i \phi \nu} - 1).
 $$
 This one dimensional integral is required for the initial condition $N=1$, and
 then the evaluation of the case $N=n$ of (\ref{A2d}) from knowledge of the
 explicit fractional power series 
 form of 
 \begin{equation}\label{A2e}
 \int_{R_{n-1}} d \theta_1 \cdots d \theta_{n-1} \, \prod_{l=1}^{n-1} e^{i \theta_l \tilde{\mu} }
 (  e^{i \theta_l} -  e^{i \phi} )^\beta
 \prod_{1 \le j < k \le n-1} (  e^{i \theta_j} - e^{i \theta_k} )^\beta.
 \end{equation}
 This in turn is deduced from knowledge of the same expansion for the
 case $N=n-1$ of (\ref{A2d}), then applying  
 the recurrence (\ref{5.1a}) with parameters compatible with
 (\ref{A2a}).
 
 For $\beta$ even it is possible to establish a direct relation between 
 $E_N^{\rm J}(0;(s,1);\lambda_1,\lambda_2,\beta)$ and
 $E_N^{\rm C}(0;(0,\phi);\beta)$. First, from the definitions we have
 \begin{multline*}
 E^{\rm J}_N(0;(s,1);0,\lambda_2,\beta) =
  E^{\rm J}_N(0;(0,1-s);0,\lambda_2,\beta)  \\
 = {1 \over J_{N,\lambda_2,0,\beta}}
  \int_{1-s}^1 dx_1 \cdots  \int_{1-s}^1 dx_N \,
  \prod_{l=1}^N x_l^{\lambda_2} \prod_{1 \le j < k \le N} | x_k - x_j|^\beta.
  \end{multline*}
  For $\beta$ even, the expansion (\ref{1.12}) is valid without any need to
  order the variables. Substituting in the above gives the formula
   \begin{equation}\label{A3a}
 E^{\rm J}_N(0;(s,1);0,\lambda_2,\beta) = {1 \over J_{N,\lambda_2,0,\beta}}
 \sum_\kappa \alpha_\kappa \prod_{l=1}^N {1 \over \kappa_l + \lambda_2 + 1}
 \Big ( 1 - (1 - s)^{\kappa_l + \lambda_2 + 1} \Big ).
  \end{equation}  
  Consider now $E_N^{\rm C}$. From the definitions,
  \begin{multline*}
  E^{\rm C}_N(0;( 0,\phi);\beta) =  E^{\rm C}_N(0;( 2 \pi - \phi, 2 \pi);\beta) \\
  = {1 \over \mathcal N_{\beta,N}} \int_0^{2 \pi - \phi} d \theta_1 \cdots  \int_0^{2 \pi - \phi} d \theta_N \,
\prod_{1 \le j < k \le N} | e^{i \theta_k} - e^{i \theta_j} |^\beta.
\end{multline*}
For $\beta$ even we can use the expansion (\ref{A2y}) without the need to
impose the ordering (\ref{A2x}). This shows
  \begin{equation}\label{A3b}
  E^{\rm C}_N(0;( 0,\phi);\beta) = {\tilde{\chi}  \over  \mathcal N_{\beta,N}} 
  \lim_{\mu \to 0} 
 \sum_\kappa \alpha_\kappa \prod_{l=1}^N {1 \over (\mu - (N-1) \beta/2 + \kappa_l }
 \Big (1 - e^{- i \phi (\mu     - (N-1) \beta/2 + \kappa_l }  \Big ),
  \end{equation}  
  where $\tilde{\chi} $ has modulus 1. Comparison of (\ref{A3a}) and (\ref{A3b})
  shows that for $\beta$ even we can map from $ E^{\rm J}_N$ to $E^{\rm C}_N$
  by setting $\lambda_2 +1= \mu - (N-1) \beta/2$ and $1 - s =
  e^{- i \phi}$, then taking the limit $\mu \to 0$ and adjusting the normalisation.
  

\begin{thebibliography}{10}

\bibitem{AGKWW14}
G.~Akemann, T.~Guhr, M.~Kieburg, R.~Wegner and T.~Wirtz,
\emph{Completing the picture for the smallest eigenvalue of real Wishart
matrices}, Phys.~Rev.~Lett. \textbf{113} (2014), 250201


\bibitem{An58}
T.W.~Anderson, \emph{An introduction to multivariate statistics}, Wiley, New
  York, 1958.
  
  \bibitem{Ao87}
K.~Aomoto, \emph{Jacobi polynomials associated with {Selberg's} integral}, SIAM
  J. Math. Analysis \textbf{18} (1987), 545--549.
  
   \bibitem{AK11} 
  K. Aomoto and M. Kita, Theory of hypergeometric functions, Springer, Tokyo, Japan (2011).
  
   \bibitem{As75} 
  R. Askey, \emph{Orthogonal Polynomials and Special Functions}, in: Regional Conference Series in Applied Math., \textbf{21}, SIAM, 1975.
  
  
    \bibitem{As80} 
  \bysame , \emph{Some basic hypergeometric extensions of integrals of Selberg and Andrews}, SIAM J. Math. Anal. \textbf{11} (1980), 938--951.
  
   \bibitem{As98} 
  \bysame, Letter to the SIAM minisymposium ``Problems and solutions in special functions",in: OP-SF NET 5.5 (Web resource), 1998
  
    \bibitem{BF97a}
T.H. Baker and P.J. Forrester, \emph{The {Calogero-Sutherland} model and
  generalized classical polynomials}, Commun. Math. Phys. \textbf{188} (1997),
  175--216.
  
  \bibitem{Bo10}
  F.~Bornemann,  On the numerical evaluation of distributions in random
  matrix theory: a review, {\it Markov Processes Relat. Fields} {\bf 16} (2010),  803--866.
  
  \bibitem{Ch16}
  M.~Chiani, \emph{Distribution of the largest root of a matrix for
  Roy's test in multivariate analysis of variance},
  J.~Mult.~Analysis \textbf{143} (2016), 467--471.
  
  
  \bibitem{Co05}
B.~Collins, \emph{Product of random projections, {J}acobi ensembles and
  universality problems arising from free probability}, Prob. Theory Rel.
  Fields \textbf{133} (2005), 315--344.
  
  \bibitem{Da72}
A.W. Davis, \emph{On the marginal distributions of the latent roots of the
  multivariable beta matrix}, Ann. Math. Statist. \textbf{43} (1972),
  1664--1669.



\bibitem{DF17}
P.~Diaconis and P.J. Forrester, \emph{Hurwitz and the origin of random matrix
  theory in mathematics}, Random Matrix Th. Appl. \textbf{6} (2017), 1730001.
  
  \bibitem{Du12}
  I.~Dumitriu, \emph{Smallest eigenvalue distribution of two classes of $\beta$-Jacobi
  ensembles}, J.~Math.~Phys. {\bf 53} (2012), 103301.
  
  \bibitem{Ed91}
  A.~Edelman, \emph{The distribution and moments of the smallest eigenvalue of
  a random matrix of Wishart type}, Lin.~Alg. Appl. {\bf 159} (1991), 55--80.  
  
  \bibitem{ES06a}
A.~Edelman and B.D. Sutton, \emph{The beta-{J}acobi matrix model, the {CS}
  decomposition, and generalized singular value problems}, Found. Comput. Math.
  \textbf{8} (2008), 259--285.

  
 \bibitem{Fo93}
P.J. Forrester, \emph{Recurrence equations for the computation of correlations in the
  $1/r^2$ quantum many body system}, J. Stat. Phys. \textbf{72} (1993), 39--50.



\bibitem{Fo06a}
\bysame, \emph{Quantum conductance problems and the Jacobi ensemble}, J.
  Phys. A \textbf{39} (2006), 6861--6870.

\bibitem{Fo10}
\bysame, \emph{Log-gases and random matrices}, Princeton University Press,
  Princeton, NJ, 2010.
  
  \bibitem{Fo12}
\bysame, \emph{Large deviation eigenvalue density for the soft edge Laguerre and Jacobi $\beta$-ensembles}, J. Phys. A: Math. Theor.
  {\bf 45} (2012), 145201.
  
  \bibitem{FH94}
P.J. Forrester and T.D. Hughes, \emph{Complex {Wishart} matrices and
  conductance in mesoscopic systems: exact results}, J. Math. Phys. \textbf{35}
  (1994), 6736--6747.
  
  \bibitem{FI10a}
P.J. Forrester and M.~Ito, \emph{Difference system for {S}elberg correlation
  integrals}, J. Phys. A: Math. Theor. \textbf{43} (2010), 175202.
  
\bibitem{FK19}
P.J. Forrester and S.~Kumar, 
\emph{Recursion scheme for the largest $\beta$-Wishart-Laguerre
eigenvalue and Landauer conductance in quantum transport}, J. Phys. A: Math. Theor. \textbf{52} (2019), 42LT02. 

\bibitem{FT19a} P. J. Forrester and A. K.~Trinh,
\emph{Optimal soft edge scaling variables for the Gaussian and Laguerre even ensembles}, Nucl. Phys. B {\bf 938} (2019), 621--639.


\bibitem{FT19} \bysame,
\emph{Finite size corrections at the hard edge for the Laguerre $\beta$ ensemble},
Stud. Appl. Math. \textbf{143} (2019), 315--336.


\bibitem{FR12} P.J. Forrester and E. M. Rains, 
\emph{A Fuchsian matrix differential equation for Selberg correlation integrals},
Commun. Math. Phys. \textbf{309} (2012), 771.


\bibitem{FW08}
P.J. Forrester and S.O. Warnaar, \emph{The importance of the {S}elberg
  integral}, Bull. Am. Math. Soc. \textbf{45} (2008), 489--534.


\bibitem{HF12} D. Holcomb and G.R.M. Flores, \emph{Edge Scaling of the $\beta$-Jacobi ensemble}, J. Stat. Phys.
\textbf{149} (2012), 1136--1160.

\bibitem{Ja75} A.T. James, {\it Special functions of matrix and single argument in statistics}, in Theory and Applications of Special Functions (R. A. Askey, Ed.), Academic, New York, 1975, pp. 497-520

 \bibitem{Jo08}
 I.M. Johnstone, \emph{Multivariate Analysis and Jacobi Ensembles: Largest Eigenvalue,
Tracy-Widom limits and rates of convergence}, Ann. Stat. \textbf{36} (2008), 2638--2716.
  
  \bibitem{JN17}
  I.M.~Johnstone and B.~Nadler, \emph{Roy's largest root test under rank-one alternatives},
  Biometrika \textbf{104} (2017), 181--193.
  
  \bibitem{Ka93}
  J.~Kaneko, \emph{Selberg integrals and hypergeometric functions associated with
  {Jack} polynomials}, SIAM J. Math Anal. \textbf{24} (1993), 1086--1110.
  
  \bibitem{Ku19} S. Kumar, \emph{Recursion for the Smallest Eigenvalue Density of beta-Wishart-Laguerre Ensemble},
  J. Stat. Phys. {\bf 175}, (2019) 126.
%

  
  
  \bibitem{MPS20}
  S.N.~Majumdar A.~Pal and G.~Schehr,
 \emph{Extreme eigenvalue statistics of correlated random variables: a pedagogical review},
  Physics Reports {\bf 840} (2020), 1--32.
  
  
 \bibitem{MS14} S.N.~Majumdar and G.~Schehr,
 Top  eigenvalue  of  a  random  matrix:  large  deviations  and  third  order  phase  transition,  J.Stat. Mech. 
 \textbf{2014} (2014) P01012 .  
 
 
 \bibitem{Me67}
M.L. Mehta, \emph{Random matrices and the statistical theory of energy levels},
  Academic Press, New York, 1967.
  
  \bibitem{Me74} 
 \bysame, Problem 74--6, Three multiple integrals, SIAM Rev. \textbf{16} (1974), 256--257
 
 \bibitem{MD63}
M.L. Mehta and F.J. Dyson, \emph{Statistical theory of the energy levels of
  complex systems. {V}}, J. Math. Phys. \textbf{4} (1963), 713--719.



 
 \bibitem{MMM19} L.~Moreno-Pozas, D.~Morales-Jimenez and M.R.~McKay,
 \emph{Extreme eigenvalue distributions of Jacobi ensembles: new exact
 representations, asymptotics and finite size corrections}, Nucl.~Phys.~B {\bf 947} (2019), 114724.
 
  \bibitem{Mo82}
 W.G. Morris, \emph{Constant Term Identities for Finite and Affine Root Systems: Conjectures and Theorems}, Ph.D. thesis, Univ. Wisconsin--Madison, 1982.


\bibitem{Mu82}
R.J. Muirhead, \emph{Aspects of multivariate statistical theory}, Wiley, New
  York, 1982.

\bibitem{NF98} T.~Nagao and P.J. Forrester, \emph{The smallest eigenvalue distribution at the
spectrum edge of random matrices}, Nucl.~Phys. B {\bf 509} (1998), 561--598.


\bibitem{Os18}
D. Ostrovsky,  \emph{A Review of conjectured laws of total mass of Bacry-Muzy GMC measures on the interval and circle and their applications},  Rev. Math. Phys. {\bf 30}  (2018), 1830003.

\bibitem{Ru16} I.~Rumanov,
\emph{Painlev\'e representation of Tracy-Widom${}_\beta$ distribution for $\beta = 6$}, Commun. Math. Phys.
\textbf{342} (2016), 843--868.

\bibitem{Se44}
A.~Selberg, \emph{Bemerkninger om et multipelt integral}, Norsk. Mat. Tidsskr.
  \textbf{24} (1944), 71--78.


\bibitem{TW94c}
C.A. Tracy and H.~Widom, \emph{Fredholm determinants, differential equations
  and matrix models}, Commun. Math. Phys. \textbf{163} (1994), 33--72. 


\bibitem{Mtmk}
Wolfram Research Inc. Mathematica Version 12 (2019).

\end{thebibliography}
\nopagebreak

\providecommand{\bysame}{\leavevmode\hbox to3em{\hrulefill}\thinspace}
\providecommand{\MR}{\relax\ifhmode\unskip\space\fi MR }
\providecommand{\MRhref}[2]{%
  \href{http://www.ams.org/mathscinet-getitem?mr=#1}{#2}
}
\providecommand{\href}[2]{#2}

\end{document}